\documentclass[preprint]{elsarticle}
\usepackage{lineno,hyperref}
\modulolinenumbers[5]
\journal{}
\usepackage{setspace}

\usepackage{geometry}

\newgeometry{vmargin={25mm}, hmargin={20mm,20mm}}
\usepackage{color}
\usepackage{amsmath}

\begin{document}

\begin{frontmatter}

\title{Dynamic tracking with model-based forecasting for the spread of the COVID-19 pandemic}

\author[IC_mainaddress]{Ian Cooper}

\author[AM_CA_mainaddress]{Argha Mondal\corref{AMcorrespondingauthor}}
\cortext[AMcorrespondingauthor]{Corresponding author}
\ead{arghamondalb1@gmail.com}

\author[AM_CA_mainaddress]{Chris G. Antonopoulos}
%\cortext[CAcorrespondingauthor]{Corresponding author}
%\ead{canton@essex.ac.uk}

\address[IC_mainaddress]{School of Physics, The University of Sydney, Sydney, Australia}
\address[AM_CA_mainaddress]{Department of Mathematical Sciences, University of Essex, Wivenhoe Park, UK}

\begin{abstract}
In this paper, a susceptible-infected-removed (SIR) model has been used to track the evolution of the spread of the COVID-19 virus in four countries of interest. In particular, the epidemic model, that depends on some basic characteristics, has been applied to model the time evolution of the disease in Italy, India, South Korea and Iran. The economic, social and health consequences of the  spread of the virus have been cataclysmic. Hence, it is essential that available mathematical models can be developed and used for the comparison to be made between published data sets and model predictions. The predictions estimated from the SIR model here, can be used in both the qualitative and quantitative analysis of the spread. It gives an insight into the spread of the virus that the published data alone cannot do by updating them and the model on a daily basis. For example,  it is possible to detect the early onset of a spike in infections or the development of a second wave using our modeling approach. We considered data from March to June, 2020, when different communities are severely affected. We demonstrate predictions depending on the model's parameters related to the spread of COVID-19 until September 2020. By comparing the published data and model results, we conclude that in this way, it may be possible to better reflect the success or failure of the adequate measures implemented by governments and individuals to mitigate and control the current pandemic.
\end{abstract}


\end{frontmatter}

%\linenumbers

\section{Introduction}

The COVID-19 pandemic becomes most significant and devastating health threat in different countries around the globe \cite{Who2019,Bocca2020,Wu2020,World2020,Remu2020}. Unfortunately, the death rates have been increasing in an unprecedented way. The disease has serious impact on the society and social consequences. It has been shown that millions of people all over the world have been extremely affected due to the spread of the virus \cite{Sohr2020,LiuY2020,Ander2020}. The outbreak of the novel coronavirus disease has been declared a pandemic by the world health organization (WHO) on March 11, 2020 and named as COVID-19 \cite{Who2019}. The novel strand of Coronavirus was first identified in Wuhan, Hubei Province, China in December, 2019 causing a severe and potentially fatal respiratory syndrome known as severe acute respiratory syndrome coronavirus 2 (SARS-CoV-2) \cite{Kraemer2020,Fan2020,Xue2020}. After more than six months, despite various steps to stop the spread of the virus by governments, by June, 2020 most countries have been affected and the total infections around the globe has exceeded nine million and the number of deaths is approaching half-million. In the absence of a proper vaccine or medicine, from the last couple of months, self-quarantine, social distancing, frequent hand washing, and wearing an antiviral face mask have  emerged as the most widely-used strategies for the mitigation and control of the current pandemic \cite{Giordano2020,Kucha2020,Liv2020,Scarpino2019,Chinazzi2020}.

Some countries have been more successful in controling it for an extended period than others. Thus, it is imperative that mathematical models can be used to monitor the spread of the virus to give a scientific basis for the control measures implemented by governments and to assess their success \cite{Ndai2020,Li2020,Anas2020,Post2020}. Such models can be used to predict the number of infections and deaths in the near future, to monitor any changes in the trends of the spread of the infection and provide estimates of time scales involved. Using mathematical models, one can gain a better quantitative understanding of the spread and control of the virus as well as provide a theoretical framework to analyse the published data for the spread of the disease. Interestingly, a second wave of infections can develop during a pandemic where the disease infects one group of people first, then, the infections appear to decrease,  but it is then followed by an increasing number of following-up infections. By comparing published data with model predictions, it is possible to detect this spike in infections early on and to judge the severity of the consequences if the upward trend in infections continues. 
Such insights can be drawn from the mathematical models that may be nearly impossible to discern from the data alone.

Researchers have been working on the fast estimations of the outbreak dynamics, its impact on people, some possible required measures in the health system, societies and so on. A suitable model can be used for estimates, using the up-to-date  data sets of COVID-19 cases in various communities.
Recently, there has been a number of published papers proposed to achieve the aim from modeling approaches using different parameters and various perspectives in different communities \cite{Fan2020,Toda2020,Peng2020,Wang2020,Shim2020,Liu2020,Singh2020,Ranjan2020,Lancet2020}. Among the epidemic models, the most notable is the SIR model \cite{Post2020,Heth1989,Heth2000,Heth2009,Weiss2013}. It is a popular model in disease modeling that is subdivided into three categories, i.e., Susceptible $S$, Infected $I$, and Removed $R$ populations, that interact in time $t$ according to the kinetic method \cite{Fan2020,Post2020}. This type of model may be appropriate to estimate the dynamics of the COVID-19 outbreak and it is based on available, published data sets, for example in \cite{Corona}. With the SIR model, the data and model parameters can be updated on a daily basis to better estimate the progress of the virus spreading within a community. An infected individual interacts with other system variables and transmits the disease with a certain rate if the other individual is susceptible. An infected individual may recover or die at a certain rate. The model dynamics can be described as a system of coupled, nonlinear ordinary differential equations (ODEs). The dynamical behavior is completely determined by the rates of the three variables with particular initial conditions. Here, we derive the conditions when an epidemic occurs and characterize its peak values for the data published in \cite{Corona} for Italy, India, South Korea and Iran.

The standard SIR model assumes a homogeneous mixing of infected individuals and a constant total population, with the susceptible population decreasing monotonically towards zero. These assumptions may not be realistic, though! The standard SIR model may not be applied successfully to study the evolution of the virus in countries such as India and Iran  for example, that we study here, where surges occur in time. Here, we adopt a revised approach that allows for the inclusion of surges in the number of infected people. The improved explanations of the SIR model adopted here do not consider the total population, or a homogenous mixing of infected individuals within a community. Rather, as people become infected and move about, more individuals are added to the susceptible population. Hence, the susceptible population is considered as a variable that can be increased at various times to account for newly infected individuals spreading throughout a community \cite{Weiss2013,Cooper2020}, i.e., to surges. The SIR model in different forms have been used in some previous studies to investigate the spread of COVID-19 \cite{Fan2020,Post2020,Toda2020}. Here, we show that our approach of the combination of the standard SIR model and the inclusion of surges, allows for the detection of the early onset of spikes in infections or the development of second waves.

We used COVID-19 data sets \cite{Corona} in the form of time-series from early to mid 2020 with various starting dates for Italy, India, South Korea and Iran. These  data sets can be compared  with the predictions of the SIR model used here. By doing so, one can make more informed decisions about the trajectories of the virus in these countries. There are also some limitation to consider. The  data sets published may not be complete. Also, countries report their data using different ways of counting. Visual inspection of plots of the  data sets against model predictions is the basis for choosing values of the input parameters in the model. Although this is very subjective, it may be a better approach than using an automatic optimization program for calculating the input parameters. The SIR model has previously been applied to Italy, India, South Korea, USA, China, Australia and the US state of Texas \cite{Cooper2020}. Here, by manually adjusting the model's input parameters, the model's predictions have been able to track the trajectory of the cumulative total number of infections, the current active number of cases, the number of recoveries and deaths reasonably well by visual inspections, for all countries considered.

The countries selected for this study have very different trajectories for the evolution of the number of infections. In Italy, the peak in the number of active cases occurred around 19 April, 2020 and this number has steadily declined ever since \cite{World2020,Remu2020,Fan2020,Giordano2020}. Unfortunately, important factors such as population density, insufficient evidence of various symptoms and transmission mechanism, make it difficult to deal with such a pandemic, especially in a high population density country such as India \cite{Ranjan2020,MOH2020}. The virus is out of control in India, and consequently, the number of infections and deaths is now increasing at an alarming rate. South Korea is acknowledged as one of the countries that have been most successful in controlling the spread of the disease, however, in early June, the number of infections has started to increase again. By using the SIR model, we can estimate how serious this event may be. In the beginning of May, 2020, a second wave has developed in Iran. The first peak in active cases occurred in early April, 2020 and the second peak with similar height has occurred in June, 2020. Using the SIR model, these events can be tracked and we can get a glimpse of what the future may hold, especially at this time when many governments and people are thinking that the virus is over and want things to get back to normal. The main goal of this paper is to compare the results between epidemiological modeling theory and actual data-driven results when fitting the model with data sets. Our approach and results provide a simple procedure to obtain the different dynamics relevant to control the spread of the disease in the studied countries, as well as in any other country or community.

The paper is organized as following: In Sec. \ref{sec_SIR_model}, we introduce the SIR model description with different populations and discuss its aspects. In Sec. \ref{sec_methodology_and_results}, we analyze the approach that we used to study the published data sets with our modeling approach and in Sec. \ref{sec_results}, we show the results of our analysis for Italy, India, South Korea, and Iran. Finally, in Sec. \ref{sec_conclusions}, we conclude our work and report the outcomes of the results and its relation with evidence already collected on the spread of COVID-19 in these countries.

\section{The standard Susceptible-Infected-Removed (SIR) model}\label{sec_SIR_model}

In this work, we consider the standard  SIR model given by the system of three coupled non-linear ODEs \eqref{SIR_model_ODEs} \cite{Toda2020,Weiss2013,Cooper2020,Amaro2020}, describing the evolution of the susceptible $S$, infected $I$ and removed $R$ populations in time $t$. It can be easily implemented to obtain a better understanding of how the COVID-19 virus spreads over time within communities, including the possibility of surges in the susceptible population. Thus, the model is designed to remove many complexities in such a way that becomes useful both quantitatively and qualitatively to estimate different outcomes. In particular, it is represented by an autonomous, continuous dynamical system that describes the time evolution of the following three populations
\begin{enumerate}
\item {{\it Susceptible individuals}, $S(t)$: The particular individuals who are not infected, however, could become infected. A susceptible individual may become infected or remain susceptible. As the virus spreads from its origin or new sources occur, more individuals will become infected, thus the susceptible population will increase for a period of time (surge period).}
\item{{\it Infected individuals}, $I(t)$: Those individuals who have already been infected by the virus and can transmit it to other individuals who are susceptible. An infected individual may remain infected, and can be removed from the infected population to recover or die.}
\item{{\it Removed individuals}, $R(t)$: These are those individuals who have recovered from the virus and are assumed to be immune from being infected again, $C(t)$ or have died, $D(t)$.}
\end{enumerate}

The time evolution of the three populations (based on the above assumptions and concepts) is governed by the following system of ODEs
\begin{equation}\label{SIR_model_ODEs}
\begin{aligned}
\frac{dS(t)}{dt}&=-aS(t)I(t),\\
\frac{dI(t)}{dt}&=aS(t)I(t) - bI(t),\\
\frac{d{R}(t)}{dt}&=bI(t),
\end{aligned}
\end{equation}
where, $a$ is the transmission rate constant and $b$ the removal rate constant. The SIR, disease transmission, model is derived assuming several strong assumptions. In particular, it is assumed that the members of the susceptible and infected populations are homogeneously distributed in space and time. An individual removed from the infected population has a lifetime immunity. The total population $N$ is constant in time, where $N = S + I + R$ and the number of births and deaths from causes other than the virus are ignored. The system of equations \eqref{SIR_model_ODEs}) can be solved using Euler or the Runge-Kutta methods.
\begin{figure}
\centering
\includegraphics[width=14cm,height=8.5cm]{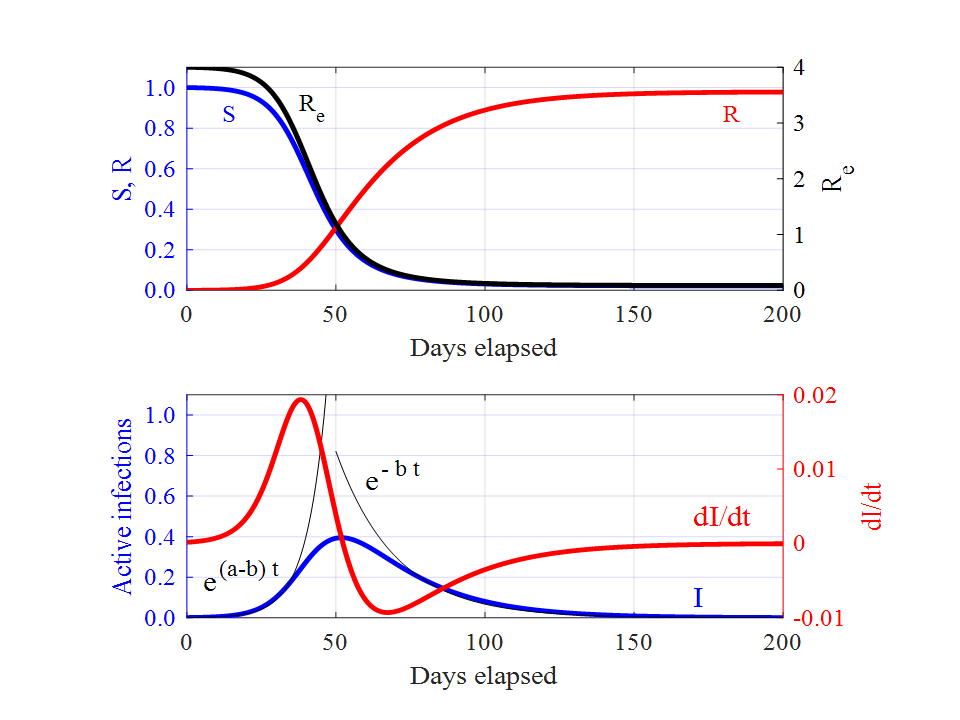}
\caption{SIR model: The time variation of the population: susceptible $S$, infected $I$, removed $R$, and effective reproductive number, $R_e$. The initial conditions and model parameters considered here are: $S(0) = 1.0$, $I(0) = 0.001$, $R(0) = 0$, $a = 0.20$, $b = 0.05$.}\label{Fig1}
\end{figure}

The evolution of the infected population is governed by the second ODE in system \eqref{SIR_model_ODEs}. In the beginning of the spread of an epidemic, where $S \approx 1$, the number of infections increases exponentially $I = I(0)\,{e^{(a - b)\,t}}$. Then, the rate of infection falls to zero at the peak where $dI/dt=0$. When $S$ drops below about $0.2$, the infected population decreases exponentially as $I \propto \,{e^{- b\,t}}$. A set of solutions is shown in Fig. \ref{Fig1} for a particular choice of initial conditions and model parameters, where one can see how the susceptible $S$, infected $I$, removed $R$, and effective reproductive number $R_e$, evolve in time. Next, we discuss the importance of $R_e$ in the context of virus spread.

We can define an effective reproductive rate $R_e$, where $R_e = a S / b$. This is important as the evolution of the disease depends upon the value of this rate. When $R_e$ becomes less than one, the infected population decreases monotonically to zero, whereas when greater than one, it increases. Crucially, at $R_e=1$, the rate of increase of the infected population is zero, what corresponds to the peak number of active infected individuals. Thus, there is a critical value for the susceptible population, $S_C = b/a$. The number of active cases only declines when $S<S_C$. The reproductive rate $R_e$ and critical susceptibility $S_C$ act as a threshold value that determines whether an infectious disease will quickly die out or continue to grow. As the existence of a threshold for infection is not obvious from the data, it can be discerned from the properties of the SIR model \eqref{SIR_model_ODEs}, what makes it a powerful tool in understanding and predicting the spread of an epidemic, such the one due to COVID-19.

\begin{figure}
	\centering	\includegraphics[width=14cm,height=8.5cm]{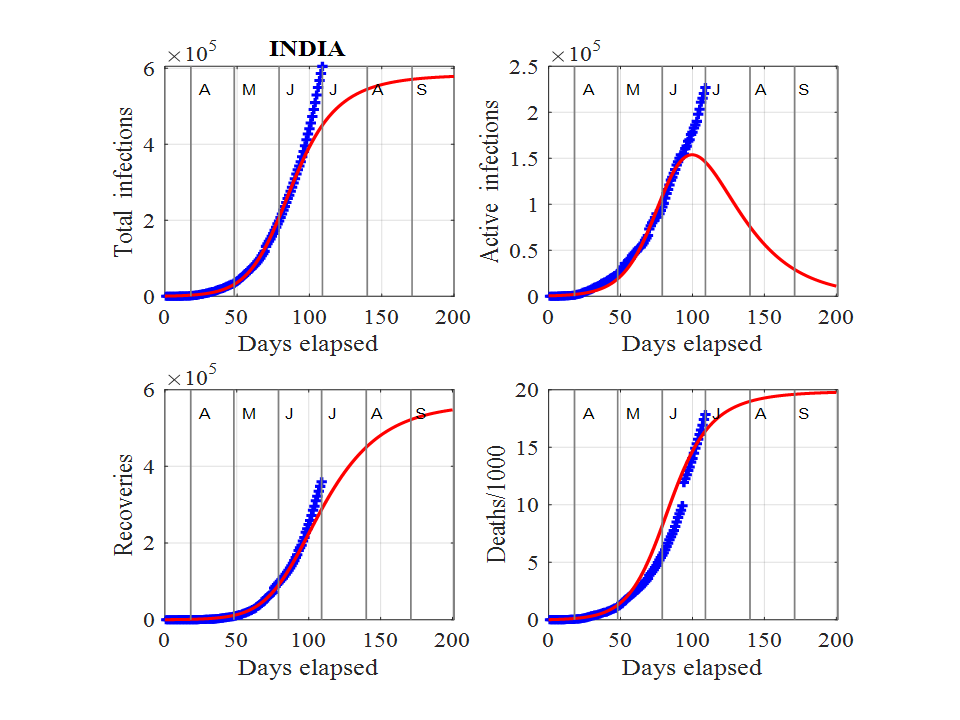}
	\caption{India: The SIR model predictions (14 March, 2020 to 30 September, 2020) and data (14 March, 2020 to 29 June, 2020) for the infections, recoveries and deaths. We note that the number of active infections in India does not reach a peak (see plot of Active infections vs days elapsed), and that it keeps increasing in time. Thus,	the solutions to the SIR model \eqref{SIR_model_ODEs} and predictions are only in agreement at the start of the pandemic in India, i.e., the period for which the red (model solutions) and blue (published data in \cite{Corona}) curves of the active infections are in good agreement.}\label{Fig2}
\end{figure}

The rate of increase in the number of infections depends on the product of the number of infected and susceptible individuals. An understanding of system \eqref{SIR_model_ODEs} explains the staggering increase in the infection rate around the world and currently, in India. The spread of the virus can only be contained when the number of susceptible individuals decreases with time. Our analysis shows that one cannot get accurate predictions by applying the SIR model to the data for the pandemic in India. Figure \ref{Fig2} identifies the problem in applying the SIR model to the Indian data starting on 14 March, 2020. The model and predictions are only in agreement at the start of the pandemic. The number of active infections in India does not reach a peak, it keeps getting bigger and bigger. However, modifying the SIR model by resetting the susceptible population at surge times, it is possible to fit the model predictions to all the available data. Therefore, applying the SIR model to the Indian data augmented by the introduction of surge times in the susceptible population, one can make a quantitative analysis due to the spread of the virus in the country.

\section{Our modeling approach}\label{sec_methodology_and_results}

In most of the countries such as India, there is not a homogenous mixing of the infected and susceptible individuals within the population. Only part of the total population will become susceptible to infected individuals. When infected individuals persist and move about within a community, further individuals may become susceptible to the disease. The susceptible population will not necessarily decrease monotonically with time. As susceptible individuals become infected, these new infections act as a source for more individuals to become susceptible. This gives rise to a positive feedback loop leading to a very rapid rise in the number of active infected cases in a surge period where the number of susceptible individuals increases instead of decreasing. Here, the SIR model \eqref{SIR_model_ODEs} is considered in such a way so that at any time $t_s$, the susceptible population can be reset to $S_s(t_s)$, accommodating for surges.

The system \eqref{SIR_model_ODEs} is solved for the variables $S$, $I$ and $R$, where $S$, $R$ and $I \in [0, 1]$. These populations are multiplied by a scaling factor $f$ to give the number of individuals in each population group to match the model outputs with the published COVID-19 data sets in \cite{Corona}. The solution gives only the values for the removed population $R$ as a function of time $t$. Individuals removed from the infected population $R$ have either recovered, $C$, or have died, $D$, where $R = C+D$. The number of deaths, $D$ is estimated by fitting a function in the plot of the data for the number of actual deaths versus the data for the number of removals. A non-linear function used to find the number of deaths, $D$ from the number of removals $R$ is given by
\begin{equation}\label{nonlinear_fitting_function}
D={D_0}\,\left({1-{e^{-k_0R}}}\right),
\end{equation}
where $D_0$ and $k_0$ are constants that are suitably selected to give the best-fit between the function and the data, with an example shown in Fig. \ref{Fig4} for data from Italy. The number of recoveries, $C$ is then simply $C = R - D$. 

The model input parameters are the population scaling factor, $f$, the initial infected population, $I(0)$, the initial removed population, $R(0)$, the rate constants $a$ and $b$, death constants, $D_0$ and $k_0$, and the surge periods $t_s$ and $S_s$. This set of constants is appropriately considered to best-fit the model predictions with the data sets \cite{Corona} for each country in our study.  As new data become available, a small amount of adjustment to the input parameters in model \eqref{SIR_model_ODEs} can be made to get a better agreement between the model predictions and the data for each population. The model cannot predict the time and height of the peak in the number of active infections. However, if the number of active infections keeps increasing, the variable $S$ can be increased which has the effect on increasing the peak value for active infections and delaying the time at which the peak value will occur. The adjustments that are necessary in resetting the value of $S$ are an important indicator of the success or failure of government actions in controlling the spread of the virus. High values of resetting of $S$ and a high reset frequency indicates that the spread of the virus is not contained in the community. Only when the susceptible population decreases towards zero, the number of active infections also declines to zero. Although the peak for active infections has not been reached in India, the model augmented by the introduction of surges as proposed here, gives the estimates of the duration of the pandemic and minimum estimates for the total cumulative number of infections and deaths, what allows for valuable forecasts.

\section{Results}\label{sec_results}

Data have been collected from \cite{Corona} for Italy, India, South Korea and Iran until the end of June, 2020. The system of Eqs. \eqref{SIR_model_ODEs} was solved by using the fourth order Runge-Kutta method. The model predictions and data for each country are displayed graphically for a period of 200 days. The crosses are for the data and the smooth lines for the model predictions from \eqref{SIR_model_ODEs}. The adjustable model input parameters are: the population scaling factor, $f$, the initial number of infections, $I(0)$, the initial number of removals $R(0)$, the transmission rate constant, $a$, the removal rate constant, $b$, and the constants $k_0$ and $D_0$ to calculate the number of deaths from the number of removals $R$. The results of our study are then shown in Figs. \ref{Fig2} to \ref{Fig13}. The time evolution figures are the cumulative total number of infections, $I_{tot}$, the active cases, $I$, recoveries, $C$, deaths, $D$ and susceptible populations, $S$ (not scaled, the horizontal line gives the critical value, $S_C$ and the dot gives the current value of $S$ for the last day in the data). The figures of the deaths, $D$ against removals, $R$ are used to find a function that best describes the relationship between $D$ and $R$. The figures of removals, $R$ against total infection, $I_{tot}$ are useful for fitting the input parameters and may be useful as indicators of a change in trend or the peak in the number of active cases. The plots of removals, $R$ against the number of active infections, $I$ are also useful for fitting the input parameters, change in trend and identifying the peak in active infections. The input parameters are adjusted by visual inspection to incorporate suitable values corresponding to the best fit of the model predictions to the published data. As we have checked, it may be possible to use an optimization program to fit the parameters, however, we found out that it will not necessarily improve the predictions in tracking the data due to the quality of the recorded data.

We start by analysing the data from Italy and then move on to the studies of the data sets from India, South Korea, and Iran.

\subsection{Italy}

\begin{figure}
\centering
\includegraphics[width=14cm,height=8.5cm]{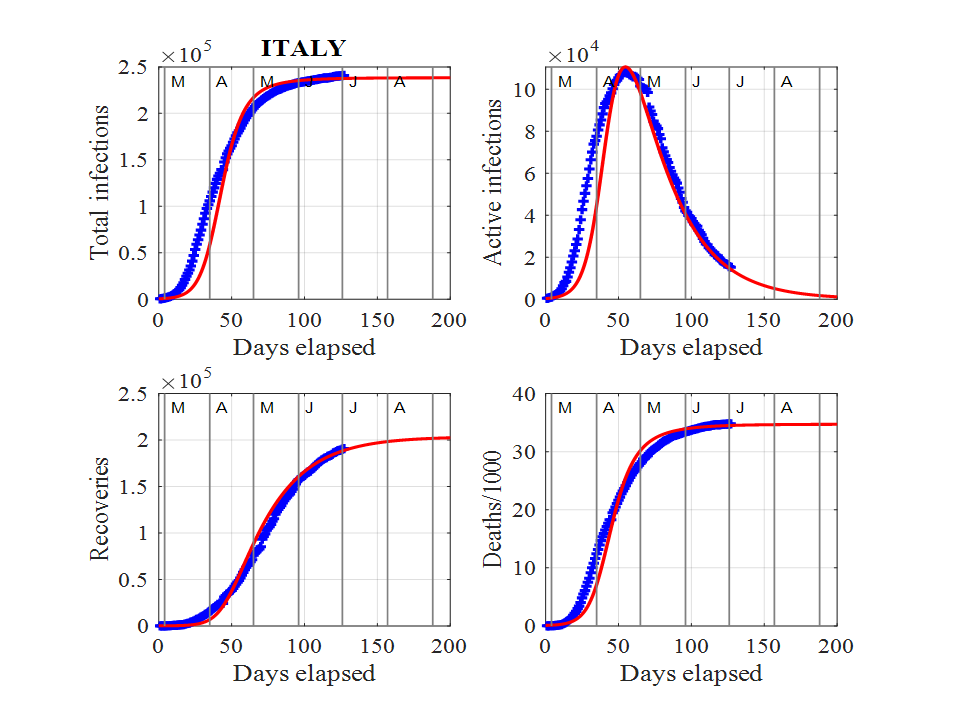}
\caption{Italy: Time evolution of the populations of cumulative total infections, active infections, recoveries and deaths/1000. Input parameters used in the modeling: $I(0) = 1.30 \times 10^{-3}$, $R(0) = 6.21 \times 10^{-4}$, $f = 2.4\times 10^5$, $a = 0.180$, $b = 0.037$, $D_0 = 3.55 \times 10^4$ and $k_0 = 1.6 \times 10^{-5}$. We note the peak in the plot of active infections vs days elapsed.}\label{Fig3}
\end{figure}
% % % % % % % % % % % % % % % % % % % % % % % % % % %
\begin{figure}
\centering
\includegraphics[width=14cm,height=8.5cm]{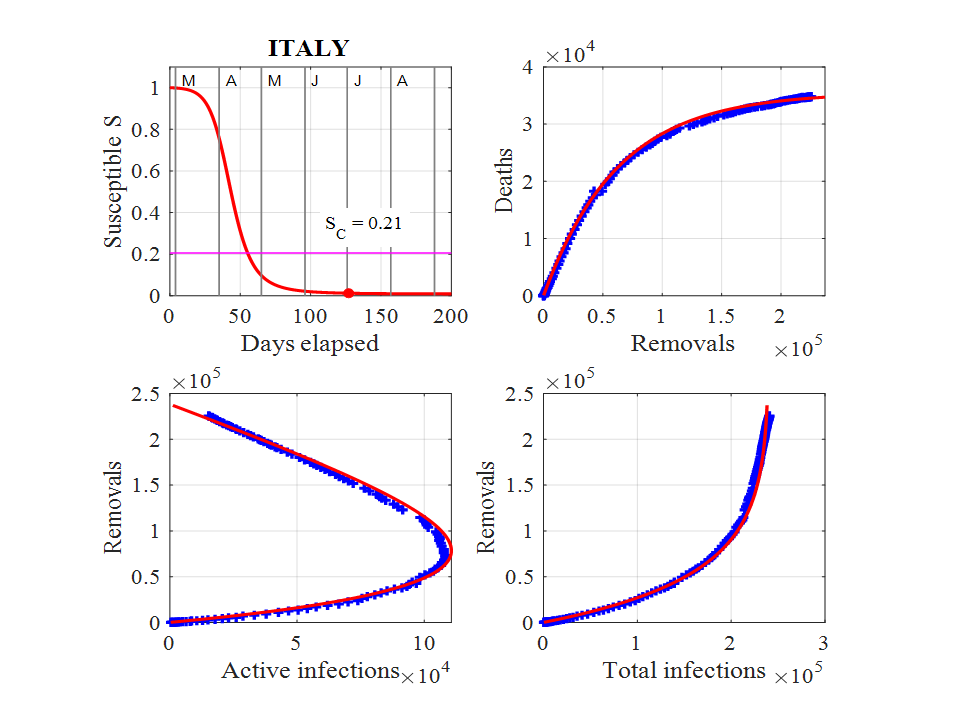}
\caption{Italy: Susceptible population as a function of time, deaths against removals, removals against active infections and removals against cumulative total infections, where $S_c$ is the critical value for the susceptible population $S$.}\label{Fig4}
\end{figure}

Figures \ref{Fig3} and \ref{Fig4} show our modelling results for the data from Italy. The virus has left a trail of suffering throughout the country, especially in the northern regions. Italy was the first country in Europe to suffer severe effects of the virus \cite{World2020,Remu2020,Fan2020,Giordano2020,Liv2020}. The numbers of infections and deaths increased very rapidly from mid-March until the end of April, 2020. This led to hospitals being quickly overwhelmed and the government imposed a strict lockdown for two months. The lockdown period was very successful with the peak infections of about 108 000 occurring around 20 April, 2020 and the susceptible population falling below the critical value. Since the peak, the number of active infections has consistently decreased as predicted by the model. In all figures for Italy, our approach has been able to track the published data exceedingly well. Also, the relationship between deaths and removals is very well described by Eq. \eqref{nonlinear_fitting_function}. This initial sharp rise in deaths and then taping off in the number of deaths as the number of removals increases, is a very common trend observed in other countries studied herein.

In early May, 2020, people were allowed to move within their regions to visit family provided they wear face masks, and some parks were reopened for exercise. Slowly, restrictions are being lifted in the country. Manufacturers and construction firms have resumed work and shops are starting reopening. There has been no need to apply any surges in the susceptible population and this is a strong indicator that so far the lifting of the restrictions has not faulted the downward spiral in the number of active infections. The model predicts that by 13 September, 2020, there will be about 35 000 deaths and about 240 000 individuals affected by the virus in the country. The numbers of infections and deaths caused by the pandemic have been huge in Italy, however, by the end of June, the actions taken by the government have succeeded in controlling the spread of the virus and infections are seen to continuously decrease.

We can explain the reasons for the large number of infections and deaths in Italy with insights gained from the SIR model \eqref{nonlinear_fitting_function}. There is strong evidence that in December, 2019 people were already infected, although the first reported case of an infected individual was not until 31 January, 2020. The rate of change of the infections depends upon the number of infections and the number of susceptible individuals as given by system \eqref{SIR_model_ODEs}. During December and January, 2020, the infected people, many of them asymptomatic, lead to an increase in the susceptible population. So, from a starting point with low number, the number of infections ballooned very rapidly, following initially an exponential increase. The recorded number of active cases on 29 February, 2002 was 1049, 3 March, 2020 was 2262 and 10578 on 11 March, 2020. Governments and their advisors around the world often stated no urgent actions are required now as the active case numbers are low, but this is the time when actions need to be taken as the initial increase of the spread follows an exponential increase, as seen in model \eqref{SIR_model_ODEs}! The Italian government was not fully aware of the presence of the virus early in 2020, hence did not take the necessary control measures to control the spread of the pandemic. For example, compare the spread of the virus in Italy and South Korea. The South Korean government took immediate actions in January and February, 2020 to control the virus and near the end of June, there has been only 280 deaths and less than 13 000 total infections. This is probably due to the fact that South Korea has a transparent and open system of government that has managed the crisis with relatively few deaths and infections. Authorities took a proactive approach to testing, contact tracing, treatment and isolation, expedited by its world-leading technological capabilities and public healthcare system.

% % % % % % % % % % % % % % % % % % % % % % % % % % %

\subsection{India}\label{results_India}

\begin{figure}
\centering
\includegraphics[width=14cm,height=7.8cm]{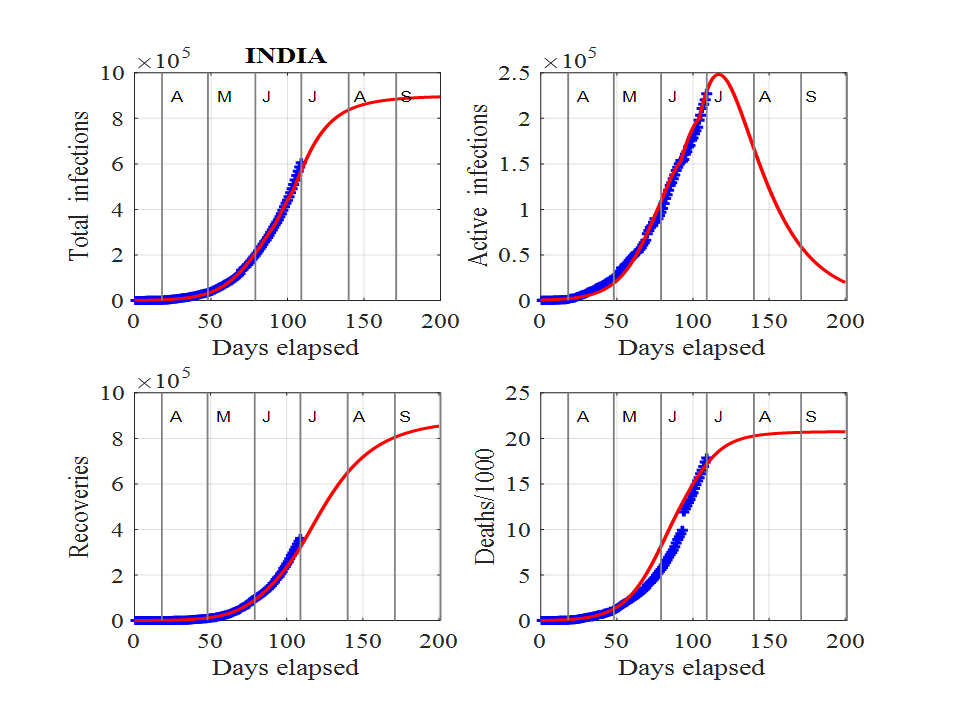}
\caption{India: Time evolution of the populations of cumulative total infections, active infections, recoveries and deaths/1000. Input parameters: $I(0) = 8.0 \times 10^{-4}, R(0) = 2.0 \times 10^{-5}, f = 6.5 \times 10^5, a = 0.122, b = 0.048$   $ D_0 = 2.0 \times 10^4$ and $ k_0 = 5.0 \times 10^{-6}$. Note the absence of a peak in the plot of active infections!}\label{Fig5}
\end{figure}
% % % % % % % % % % % % % % % % % % % % % % % % % % %
\begin{figure}
\centering
\includegraphics[width=14cm,height=8.5cm]{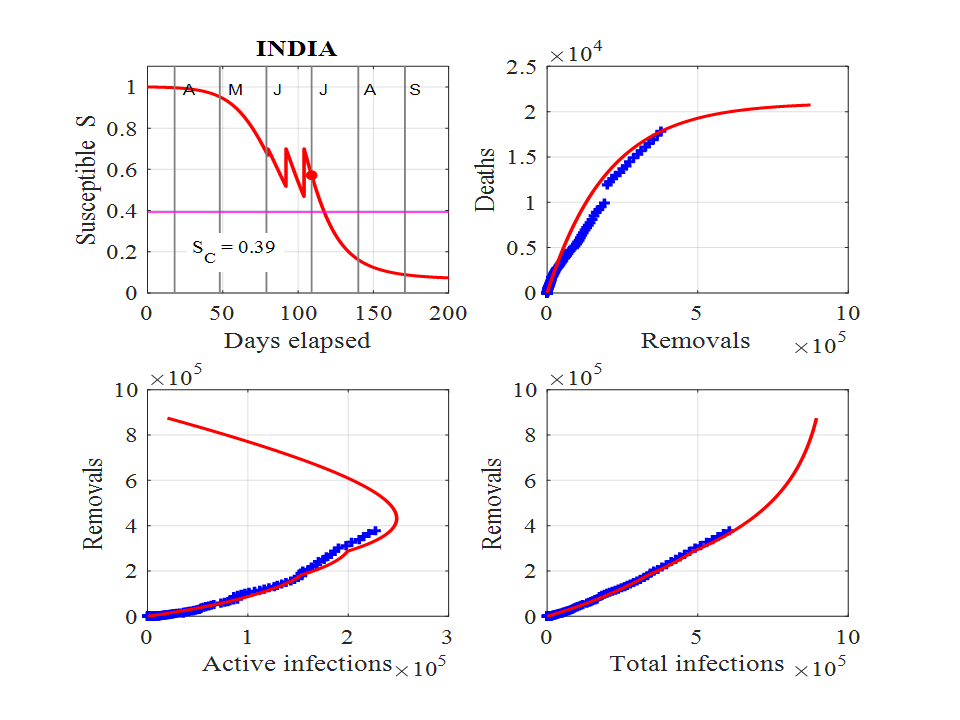}
\caption{India: Susceptible population as a function of time, deaths against removals, removals against active infections and removals against cumulative total infections, where $S_c$ is the critical value for the susceptible population $S$. The number of active infections keeps increasing because $S>S_C$ until July, 2020. }\label{Fig6}
\end{figure}
% % % % % % % % % % % % % % % % % % % % % % % % % % % % %

Figures \ref{Fig5} and \ref{Fig6} display the estimate of the data and the model predictions \eqref{SIR_model_ODEs} for India. Clearly as the end of June approaches, the spread of the virus throughout the country is not being contained and the peak number of active cases may be a long way into the future. The infected population is still increasing and a peak in the active infections has not yet been attained, as shown in Fig. \ref{Fig5}. The model is designed so that the model's predictions track the published data. Hence, the model cannot predict the peak number of active cases in the future. However, by adding surges, the peak can be delayed and increased. Although the selection of model's input parameters and surge times are subjective, the data can be tracked quite successfully and can give estimates of the populations in the future and possible time scale for the number of active cases to drop to low values, that allows for valuable predictions.

India is caught in a vicious positive feedback loop like the USA. New infections result in an increase in the susceptible population as the disease spreads. Then, these individuals may become infected and the cycle continues. This cycle is difficult to terminate, resulting in the number of infections getting larger and larger. One of the major methods in controlling the increase of infections is to test, contact trace and quarantine, however, when there is such large numbers of infections as in India and the USA, this method is not practical.

Figure \ref{Fig6} shows that the peak in the number of active cases may not occur for some time. The observation of the susceptible variable, $S$ shows that the susceptible population has to be repeatedly reset to higher values. This is a strong indicator that the spread of the virus is not contained in India. The plots for the removals against infection or total infections show that the data has not reached the knee section of the model's prediction again, providing evidence that the peak in the active infections may not occur any time soon. It is most likely that in the near future, there will be an alarming increase in the number of new infections and deaths in India. The model predictions make it possible to consider many of the scenarios that may result. 

\subsection{South Korea}

\begin{figure}
\centering
\includegraphics[width=14cm,height=8.5cm]{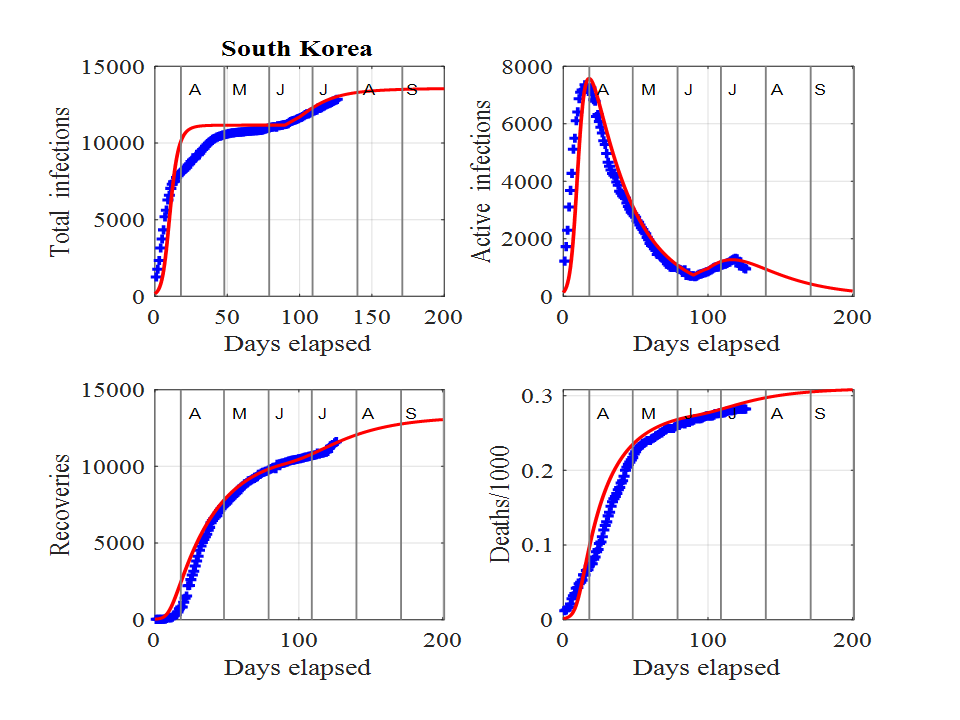}
\caption{South Korea: Time evolution of the populations of cumulative total infections, active infections, recoveries and deaths/1000. Input parameters: $I(0) = 1.2 \times 10^{-2}$, $R(0) = 3.3 \times 10^{-3}$, $f = 1.1 \times 10^4$, $a = 0.400$, $b = 0.035$, $D_0 = 400$ and $k_0 = 1.1\times 10^{-4}$. We note the two peaks in the plot of active infections vs days elapsed and that the second, lower peak corresponds to a surge.}\label{Fig7}
\end{figure}
% % % % % % % % % % % % % % % % % % % % % % % % % % %
\begin{figure}
\centering
\includegraphics[width=14cm,height=8.5cm]{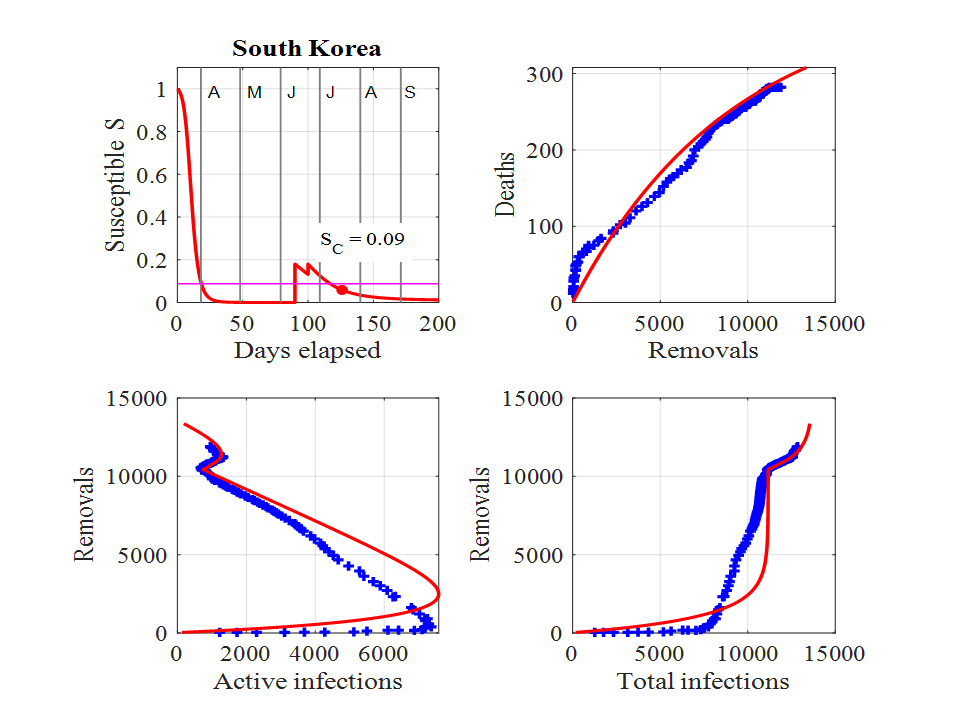}
\caption{South Korea: Susceptible population as a function of time, deaths against removals, removals against active infections and removals against cumulative total infections, where $S_c$ is the critical value for the susceptible population $S$. The application of surges produces the result $S>S_C$ in June, 2020, so that the spike in the active cases can be tracked. }\label{Fig8}
\end{figure}

South Korea has a transparent and open system of government that has managed the crisis with relatively few deaths and infections. Authorities took a proactive approach to testing, contact tracing, treatment and isolation, expedited by its world-leading technological capabilities and public healthcare system. Lives for many in South Korea have more or less rolled on as usual. It is mandatory to wear masks on public transport and most South Koreans wear masks when going out. South Korea has tried to remain as open as possible by responding proportionally to the scientific assessments of the risks.

\begin{figure}
\centering
\includegraphics[width=14cm,height=8.5cm]{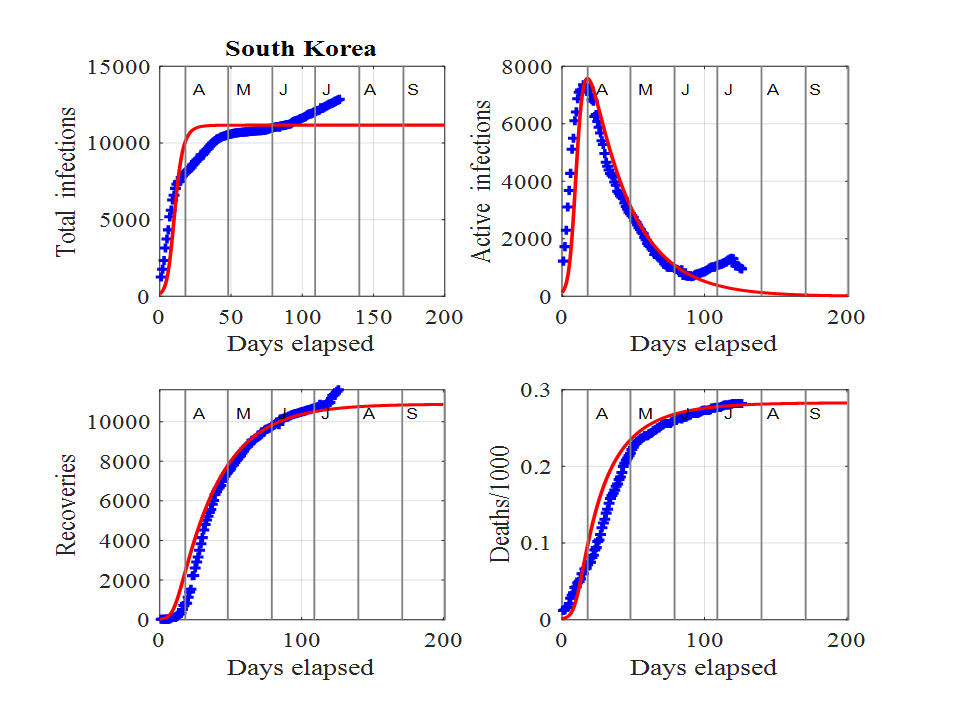}
\caption{South Korea: Time evolution of the populations of cumulative total infections, active infections, recoveries and deaths/1000 when surges have not been applied. We note the two peaks in the plot of active infections vs days elapsed.}\label{Fig9}
\end{figure}

\begin{figure}
\centering
\includegraphics[width=14cm,height=8.5cm]{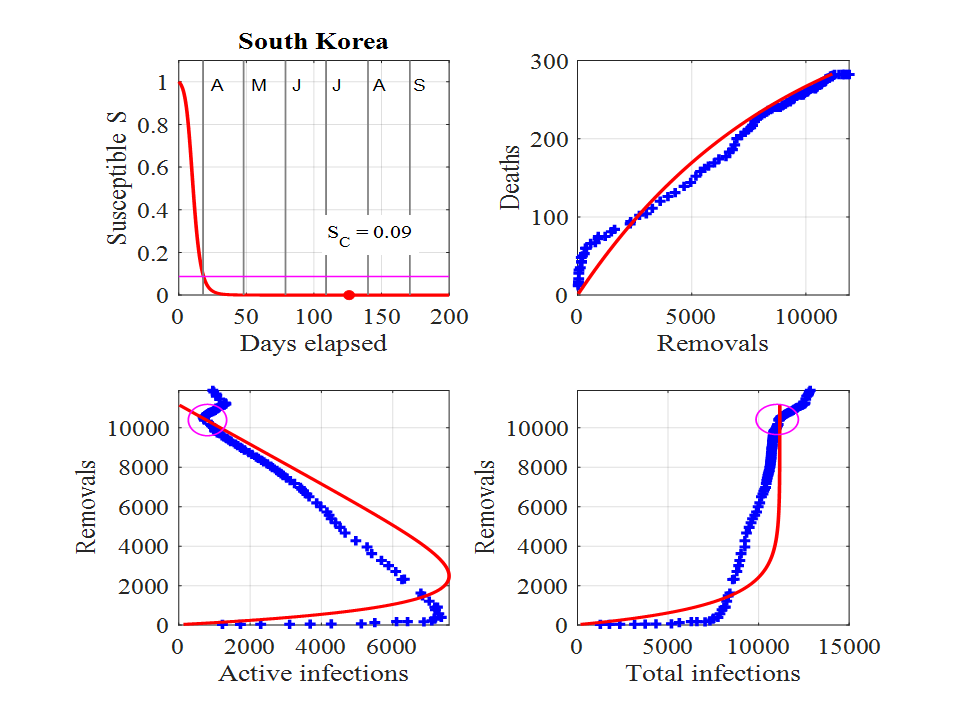}
\caption{South Korea: Susceptible population as a function of time, deaths against removals, removals against active infections and removals against cumulative total infections when surges have not been applied. The ellipses in the last two plots depict the changes in trends, which indicate the start of a surge. }\label{Fig10}
\end{figure}

However, there has been a spike in the number of infection in the early part of June, 2020. The new clusters of infections may have been the result of clubs and bars re-opening. South Korean authorities have acted swiftly by isolating some areas, testing and contact tracing and some schools have been forced to close just days after re-opening. The model \eqref{SIR_model_ODEs} can estimate this spike in the number of active cases by including surge periods where the susceptible population is increased for the model's predictions to track the data. Figures \ref{Fig7} and \ref{Fig8} show the graphical comparisons between the data and model predictions for South Korea when surge periods are included whereas Figs. \ref{Fig9} and \ref{Fig10} show the predictions if no surges were applied. Table (\ref{table1}) summarizes some of these predictions.

% % % % % % % % % % % % % % % % % % % % % % % % % % % % % % % % % % % % % % % % % % % % % % % % % % % % % % % % % % % %
% % % % % % % % % % % % % % % % % % % % % % % % % % %
\begin{figure}
\centering
\includegraphics[width=14cm,height=8.5cm]{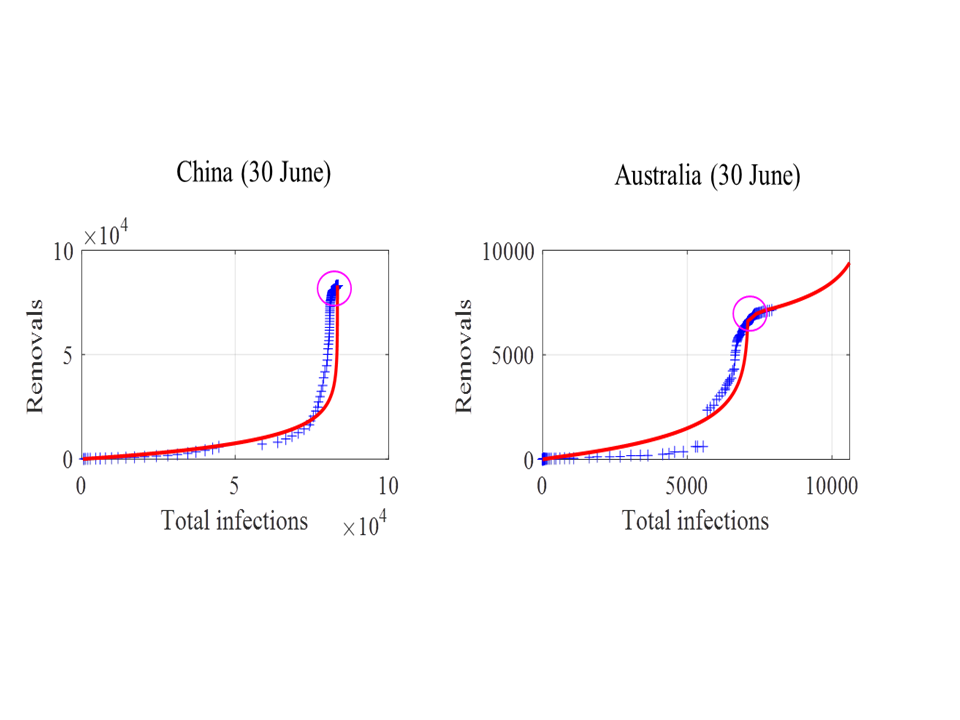}
\caption{A comparison between the removals vs infections plots for China and Australia: Plots of removals against total infections are useful in identifying a change in the trend of the course of the virus that occurred in both countries on 20 June, 2020, denoted by the ellipses.}\label{Fig11}
 \end{figure}

% % % % % % % % % % % % % % % % % % % % % % % % %
According to model \eqref{SIR_model_ODEs}, if the current trend in infections continues, then the impact of the spike in infections will not have any serious impacts. However, if these minor upward trends continue, it could lead to a very different outcome. One can identify a change of a trend for infections when there is a departure between the trajectories of the published data and model predictions. Examples of changes in trends are indicated by the ellipses which are added to the plots in Figs. \ref{Fig9} to \ref{Fig11}. Actually, the disparity between the published data and model can be used as an early warning sign of a new outset of the virus in the community. This break in downward trend in active cases is found when the model has been applied to China and Australia in June, 2020 as shown in Fig. \ref{Fig11}. In particular, it was reported on 11 June, 2020 that there has been a cluster of infections arising in  Beijing. It ended a run of 55 days without reported local transmission. Authorities have responded quickly with strong adequate measures. Another example is what happened in Victoria in Australia, where there has been a relative small spike in the number of infections as restrictions have been progressively been removed. Some of these restrictions are now being reversed. On 22 June, 2020, one million people in Melbourne have been asked not to travel outside their area and remain at home as a consequence of this spike.
\begin{table}[h!]
\centering
\begin{tabular}{|p{0.75in}|p{0.7in}|p{0.8in}|p{0.9in}|p{0.9in}|} \hline 
 		South Korea & $I_{tot}$ & \textit{I} & \textit{C} & \textit{D} \\ \hline 
 		Surges & 13779 & 197 (1.4$\% $) & 13271 (96.3$\% $) & 310 (2.3$\% $) \\ \hline 
 		No surges & 11168 & 17 (0.2$\% $) & 10868 (97.3$\% $) & 283 (2.5$\% $) \\ \hline 
\end{tabular}
\caption{South Korea: Predictions for 13 September, 2020 for the cumulative total number of infections, $I_{tot}$, active cases, $I$, recoveries, $C$ and deaths, $D$ when surges are included (second row) and no surges applied (third row). In brackets, we report the percentages of the corresponding populations.}\label{table1}
\end{table}

 % % % % % % % % % % % % % % % % % % % % % % % % % % % % % % % % % % % % % % % % % % % % % % % % % % % % % % % % % % % % % % % % % % % % % % % % % % % % % % % % % % % % % % % % % % % % % % % %
% % % % % % % % % % % % % % % % % % % % % % % % % % % % % % % % % % % % % % % % % % % % % % % % % % % % % % % % % % % % % % % % %

\subsection{Iran}

\begin{figure}
\centering
\includegraphics[width=14cm,height=8.5cm]{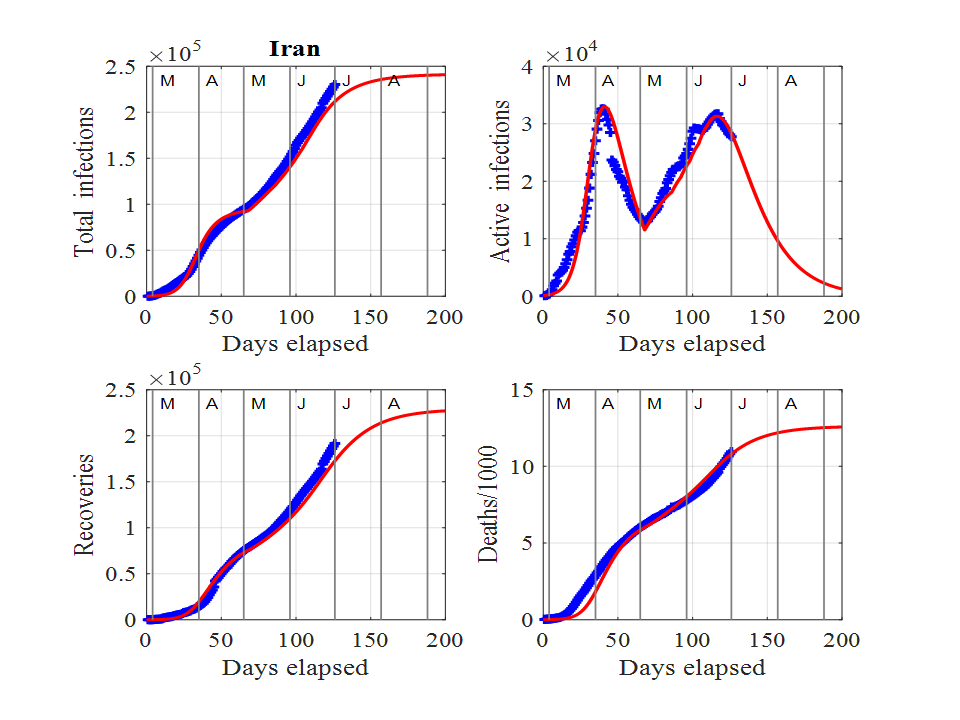}
\caption{Iran: Time evolution of the populations of cumulative total infections, active infections, recoveries and deaths/1000. Input parameters: $I(0) = 1.3 \times 10^{-3}$, $R(0) = 0$, $f = 1.0 \times 10^5$, $a = 0.250$, $b = 0.075$, $D_0 = 1.2 \times 10^4$ and $k_0 = 1.0 \times 10^{-5}$. We note the two peaks in the plot of active infections vs days elapsed, where the second corresponds to a surge. }\label{Fig12}
\end{figure}
% % % % % % % % % % % % % % % % % % % % % % % % % % %
\begin{figure}
\centering
\includegraphics[width=14cm,height=8.5cm]{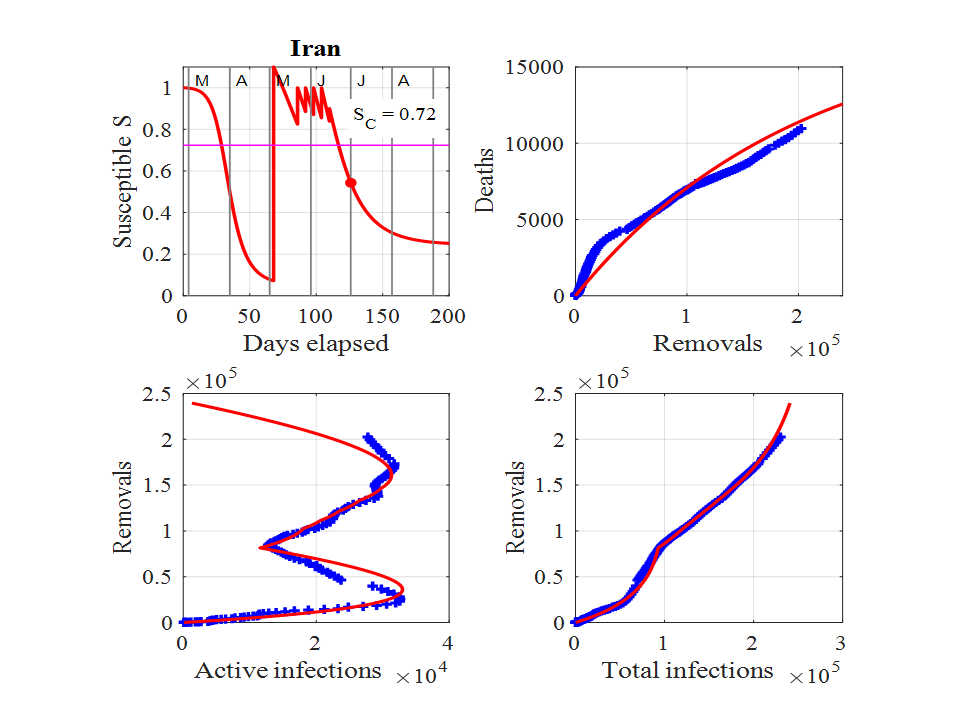}
\caption{Iran: Time evolution of the populations of cumulative total infections, active infections, recoveries and deaths/1000, where $S_c$ is the critical value for the susceptible population $S$. The surges applied in May and June, 2020, are such that $S>S_C$. While $S>S_C$, the number of active cases will continue to increase. As of 24 June, 2020, $S<S_C$ and the number of active infections has started to decline again. }\label{Fig13}
\end{figure}
% % % % % % % % % % % % % % % % % % % % % % % % % % % % % % % % % % % % % % % % % % % % % % % % % % % % % % % % %

Iran is one of the worst-hit countries in the Middle East and started easing its lockdown in April, 2020, after a drop in the number of deaths. Now, the fears that Iran is in the grip of a second wave of coronavirus have been realized. Figure \ref{Fig12} shows that a second wave of infections started in early-May, 2020 and by the end of June, 2020, the peak in the number of active cases equals the height of the initial peak in April. The second increase seems to be part of a second surge as there is a small decrease following the second peak, after 24 June, 2020. In contrast, there is no second wave (surge) for India (see subsec. \ref{results_India})) and the USA \cite{Cooper2020} as the first peak number of active cases in these countries has not yet occurred.

Government authorities in Iran have been reluctant to acknowledge that they may have lifted restrictions prematurely and in June, 2020, many more restrictions have been lifted such as the opening of mosques, shops and offices. Figures \ref{Fig12} and \ref{Fig13} highlight the dangers ahead for Iran. It is uncertain how large the second peak in active cases will be and the trend shown in Fig. \ref{Fig13}, the plot of deaths against removals shows an accelerating number of deaths given by the data, whereas the current model shows a tapering off in the number of deaths. There is a high probability that the number of deaths will far exceed 10 000. It has been reported in the public media that the actual number of deaths may exceed the reported deaths by a factor ranging from 5 to 10! The number of infections and deaths will continue to increase, unless governments and individuals take actions to prevent people becoming susceptible. To track the data, a number of surges in the susceptible population $S$ is necessary. The resetting of variable $S$ is very subjective, however, the numbers used are not as important as the fact that the susceptible population is increasing, and when this occurs, the spread of the virus is out of control and the number of infections will continually increase.

% % % % % % % % % % % % % % % % % % % % % % % % % % % % % % % % % % % % % % % % % % % % % % % % % % % % % % % % % % % % % % % %

\section{Conclusions}\label{sec_conclusions}

How the virus spreads in a community is a complex issue, affected by many factors. In this paper, the proposed approach is designed to remove many of these complexities and yet to be useful in both quantitative and qualitative ways. This is accomplished by including surge periods where the susceptible population can be reset to larger numbers allowing for the tracking of the published data sets where new clusters of infections occur in the community. The frequency at which surge periods are needed to be applied and their magnitudes, are good indicators that the spread of the virus is not under control in the community. Although the SIR model cannot predict the severity and time of the peak in the active infections, the peak in the number of active infections and the number of deaths, it does give minimum estimates for the duration of the virus, the total cumulative number of infections, the peak in the number of active infections and the number of deaths. The model and approach used here may be a useful tool in detecting the early onset of further waves of infections allowing for governments and authorities to take appropriate measures to contain them.

The virus has severely impacted all aspects of our lives around the world for more than six months now \cite{Giordano2020,Anas2020}. Typically, the half-life for the decrease in active cases is around 18 days. For India, if the peak in active cases occurred at the end of June and there were no further surges, then it would take five months for the number of active cases to drop to less than 500. Even after a peak, there is a strong probability that there will be periods when the number of infections spike again as in Iran and South Korea. Therefore, it is imperative that governments and individuals do not assume that the virus is going away and maintain or even increase measures that have been confirmed to prevent individuals becoming susceptible to the disease. Only as the susceptible population converges to zero, the number of active cases goes to zero as well.

If completely reliable data are available, the model provides an effective tool to track the spread of the virus and provide insights into the management of the control measures implemented by governments within a community. In this paper, the SIR model has been applied to data from Italy, India, South Korea and Iran. A strength of the modifications in the SIR model lies in the fact that as new data are added daily, the model parameters are easy to adjust and provide the best-fit curves between the data and the model predictions. By comparing the data with the model predictions, it is possible to reach conclusions based upon science about the effectiveness of control and adequate measures implemented by governments and people within a community.

Finally, for the virus to be controlled and to limit its impact on society, various governments have imposed drastic actions to prevent people to become susceptible to the disease. Some actions that are proving to be successful include: total lockdowns, testing, contact tracing of people who may have been in contact with an infected individual and then isolating them, isolating suspected individuals, restrictions on the movement of people, restrictions on large gatherings, hygiene measures such as washing hands, and wearing face masks particularly on public transport or in crowded indoor spaces. By using the predictions of mathematical model and comparing it with published data, one is more able to assess actions that have been taken to control the spread of the virus. Now, such advanced policies and strategies are needed to overcome the situation. Researchers from various disciplines have been working to deal with this pandemic by providing novel approaches, optimal policies, mathematical methods for predictions, biological phenomena and chemical drugs to mitigate the effects \cite{Giordano2020,Zhao2020,Rud2020,Feng2020,Bj2020}. 

\section*{Acknowledgements}
AM is thankful for the support provided by the Department of Mathematical Sciences, University of Essex, UK to complete this work.
 
%\section*{References}
 
%\bibliography{mybibfile}

\end{document}